\pdfoutput=1
\documentclass[aps,preprint]{revtex4}

\usepackage{graphicx}
\usepackage{bm}

\newcommand{\etal}      {\textit{et al.}}
\newcommand{\tc}        {$T_{c}$}
\newcommand{\lsco}      {La$_{2-x}$Sr$_x$CuO$_4$}
\newcommand{\dxxyy}     {$d_{x^2-y^2}$}

\newcommand{\dzz}       {$d_{z^2}$}
\newcommand{\psigma}    {$p_\sigma$}
\newcommand{\pz}        {$p_z$}
\newcommand{\up}        {\uparrow}
\newcommand{\down}      {\downarrow}

\begin{document}

\title{Resistance\\ of\\ High-Temperature Cuprate Superconductors}
\author{Jamil \surname{Tahir-Kheli}}
\email{jamil@caltech.edu}
\thanks{Fax: 626-525-0918. Tel: 626-395-8148.}
\affiliation{Department of Chemistry,\\ Beckman Institute (MC 139-74),
California Institute of Technology, Pasadena, CA 91125}

\begin{abstract}
Cuprate superconductors
have many different atoms per unit cell. A large
fraction of cells (5-25\%) must be modified (``doped'')  before the
material superconducts. Thus it is not surprising that there is
little consensus on the superconducting mechanism, despite
almost 200,000 papers \cite{Mann2011}.
Most astonishing is that for the simplest electrical property,
the resistance,
``despite sustained
theoretical efforts over the past two decades, its origin and its
relation to the superconducting mechanism remain a profound,
unsolved mystery \cite{Hussey2011a}.''
Currently, model parameters used to fit normal state properties
are experiment specific and vary arbitrarily from one doping to the other.
Here, we provide a
quantitative explanation for the temperature and doping dependence of
the resistivity, Hall effect, and
magnetoresistance in one self-consistent model by showing that
cuprates are intrinsically inhomogeneous with a percolating metallic
region and insulating regions.
Using simple counting of dopant-induced plaquettes,
we show that the superconducting pairing and resistivity
are due to phonons.
\end{abstract}

\maketitle

Since superconductivity requires coherent Cooper pairing of
electrons, knowing what couples most strongly to electrons
is absolutely necessary for understanding what causes
high-temperature cuprate superconductivity.
The resistivity, the Hall effect, and the magnetoresistance
are fundamentally measurements of
the momentum dependent Fermi surface scattering rate, $1/\tau(k)$,
that measures the strength of whatever is coupling to electrons.
If the origin of the scattering rate and its temperature dependence
are not understood, then most likely cuprate superconductivity is
not understood either.

The earliest resistivity ($\rho$) measurements
\cite{Gurvitch1987}
on cuprates found $\rho$ to be approximately linear in temperature
over the huge temperature range of $10<T<1000$ K.  A
linear $T$ resistivity is characteristic of electron-phonon
(or electron-boson) scattering.
Yet phonons are not believed to cause $\rho$,
despite the fact that historically,
the dominant scattering mechanisms of electrons in metals
have been phonons and impurities.
There are two reasons for this conclusion: First, at low temperatures,
the Bose-Einstein statistics of the phonons reduces the phase space
for scattering, leading to $\rho(T)\sim T^5$
(the Bloch-Gruneisen law) \cite{Ziman1972}.
The $T^5$ scaling should be
observable for $T<\Theta_{\mathrm{Debye}}/10$ where
$\Theta_{\mathrm{Debye}}$ is
the Debye temperature
(the characteristic energy of the highest energy phonons).
Since $\Theta_{\mathrm{Debye}}\sim 400$ K in cuprates
\cite{Decroux1987},
phonon scattering is not compatible with the observed linearity.
Magnons are also eliminated because
$\Theta_{\mathrm{Mag}}\sim 1500$ K
\cite{Sulewski1990}.

Second, the magnitude
of $\rho$ at high $T$ for some dopings exceeds the
Mott-Ioffe-Regel limit (MIR) \cite{Hussey2004}
that occurs when the electron mean free path
reaches the shortest
Cu$-$Cu distance ($\approx 3.8\ \mbox{\AA}$).
For cuprates,
$\rho(T)$ should saturate to
$\rho_{\mathrm{MIR}}\sim 1000\ \mu\Omega$-cm.
Instead, $\rho$ increases linearly right through $\rho_{\mathrm{MIR}}$,
leading to the conclusion that the normal state may not even
be a typical metal (a non-Fermi liquid).

Both of these conclusions are
invalid when the crystal is intrinsically inhomogeneous, as we will
show.  Briefly, the first point implicitly assumes that
phonon momentum is a good quantum number.  The
second underestimates
$\rho_{\mathrm{MIR}}$ by overestimating the density
of charge carriers.  For example, at optimal hole doping of
$x=0.16$ (we define $x$ to be the number of holes per planar CuO$_2$),
we find the fraction $4\times 0.16=0.64$ of the crystal is metallic
leading to a $\rho_{\mathrm{MIR}}$
that is $(1/0.64)=1.56$ times larger than the
conventional estimate.  $\rho(T)$ remains below
the larger $\rho_{\mathrm{MIR}}$
up to the melting temperature.

Recently, the doping and $k$-vector dependent
scattering rate, $1/\tau(x,k)$, has been extracted
from a series of beautiful experiments by
Hussey \etal\ 
\cite{Abdel-Jawad2006,Analytis2007,Hussey2011a,Cooper2009}.
In the first set of experiments
\cite{Abdel-Jawad2006,Analytis2007},
the magnetoresistance (MR) of
large single crystals of single layer Tl$_2$Ba$_2$CuO$_{6+\delta}$
was measured as a function of the direction of an applied
$45$ Tesla magnetic field for $T<40$K and large overdoping.
They found that $1/\tau(x,k)$ is the sum of three terms,
$1/\tau(x,k)=1/\tau_0(x)+A_1(x) \cos^2(2\varphi)T+A_2(x) T^2$,
where $A_1(x)$ and $A_2(x)$
are hole doping dependent constants and $\varphi$ is the angle
between the Cu$-$O bond direction and the $k$-vector as shown in the
top right corner of figure 6a.  The first term, $1/\tau_0(x)$, is a constant
that is sample dependent even at the same doping.
The anisotropic linear $T$ term is zero
for $k$-vectors along the diagonal ($\varphi=\pi/4$)
and largest for $k$ along the Cu$-$O bond directions ($\varphi=0,\pi/2$).

In an elegant theoretical analysis,
Hussey \etal\
\cite{Hussey2011a,Hussey2003,Kokalj2012}
showed that a scattering rate of the above form explained the
resistivity, the Hall effect, and the MR for a large
range of dopings, temperatures, and different cuprate materials.  The authors
suggested that the isotropic $T^2$ part of the
scattering rate arose from electron-electron Coulomb scattering while
the origin of the $T$ term is unknown.  Finally, the
strong doping dependence of $A_1(x)$ and $A_2(x)$ was noted, but not explained.

In another experiment, Hussey \etal\ \cite{Hussey2011a,Cooper2009}
used large magnetic fields to lower the \tc\ (the superconducting
transition temperature)
of \lsco. They found that $\rho(T)$ was the sum of
doping dependent $T$ and $T^2$ terms at low temperatures plus
a sample and doping dependent constant.  These results suggested that
the $1/\tau(x,k)$ form found for
Tl$_2$Ba$_2$CuO$_{6+\delta}$ was applicable to all dopings
and cuprate materials.

The authors also found that the low temperature
$\rho$ evolved into linear $T$ at high temperatures, as seen
previously \cite{Naqib2003}.
They concluded that the isotropic $T^2$ term
was due to electron-electron Coulomb scattering
and Mott-Ioffe-Regel saturation of the scattering
rate at high temperatures was reducing this term down to
linear $T$.

The experiments described above may be summarized as follows \cite{Hussey2011a}.
The temperature derivative of the resistivity,
$d\rho(x,T)/dT$, is of the form
$d\rho(x,T)/dT=\alpha_1(x,0) + 2\alpha_2(x)T$
at low temperatures and
$d\rho(x,T)/dT=\alpha_1(x,\infty)$ at high temperatures.
$\alpha_1(x,0)$ and $\alpha_1(x,\infty)$
are both doping dependent constants with
$\alpha_1(x,0)<\alpha_1(x,\infty)$ always.
$\rho$ smoothly interpolates between these two forms.
In addition,
$\alpha_1(x,0)$ is approximately constant for $x<0.19$ and
$\alpha_1(x,\infty)$ is approximately constant for $x>0.19$.
Surprisingly, the low-doping
$\alpha_1(x,0)$ constant and the high-doping $\alpha_1(x,\infty)$ constants
appear to be equal ($\approx 1\ \mu\Omega$-cm$/$K)
within the experimental error bars.
A plot of the experimental data \cite{Hussey2011a} for $\alpha_1(0)$ and
$\alpha_1(\infty)$ for \lsco\ is shown in figure 1 (for notational
convenience, we suppress the
$x$ in $\alpha_1(0)$, $\alpha_1(\infty)$, and $1/\tau(k)$
from this point onward).

Even for a fixed doping value, it is clear that
the low and high $T$ forms of $1/\tau(k)$
put strong constraints on theory.  An even stronger constraint
is explaining the doping dependence of
$\alpha_1(0)$ and $\alpha_1(\infty)$
and their peculiar ``crossover''
at doping of $x\approx 0.19$, as can be seen in figure 1.

In prior publications \cite{Tahir-Kheli2011,Tahir-Kheli2010},
we showed that cuprates are comprised of 
a percolating metallic region and insulating regions.
The metallic region is formed from 4-Cu-site square plaquettes
in the CuO$_2$ planes that may overlap (share a Cu atom).  These
plaquettes are centered around the dopants (e.g., Sr in \lsco).
The insulating regions are comprised of localized d$^9$ Cu spins
that are coupled antiferromagnetically.

By simply counting the volume and surface areas of these two regions
as a function of doping, we explained the origin
and doping dependence of the pseudogap (PG) \cite{Tahir-Kheli2011},
the universal room-temperature thermopower, the STM incommensurability,
the neutron resonance peak, and the generic cuprate phase
diagram \cite{Tahir-Kheli2010}.
Here, we show that $\alpha_1(0)$ and $\alpha_1(\infty)$
are also explained by the same intrinsic inhomogeneity (figure 2).

We will show that
$\alpha_1(0)$ is due to phonon scattering inside non-overlapping
plaquettes and $\alpha_1(\infty)$ is due to scattering from all phonons.
Therefore, $\alpha_1(0)\propto N_{4M}$ and
$\alpha_1(\infty)\propto N_{Cu}$,
where $N_{4M}$ equals the number of planar Cu sites inside non-overlapping
plaquettes and $N_{Cu}$ is the total number of planar Cu sites.
Dividing by the total number of charge carriers, $N_{M}$, leads to
$\alpha_1(0)=C(N_{4M}/N_{M})$ and $\alpha_1(\infty)=C(N_{Cu}/N_{M})$.
The best fit (see figure 1) is $C=0.904\ \mu\Omega$-cm/K.
Figure 1a has one adjustable parameter, $C$.  Figure 1b is the
ratio of $\alpha_1(0)/\alpha_1(\infty)$ and has zero adjustable
parameters.  The excellent fit in figure 1 using simple counting
is the main result of this paper.

Figure 2 shows a schematic of the distribution of plaquettes
and how $N_{4M}$, $N_{M}$, and $N_{Cu}$
are calculated.  The insulating/antiferromagnetic (AF) regions
(blue, green, and purple arrows)
and the metallic region (yellow) are shown.  The non-overlapping
4-Cu-site plaquettes are shown by black squares with the number ``$4$'' inside.
The number of Cu sites inside these squares equals $N_{4M}$.
$N_{M}$ is the number of Cu
sites in the metal region (yellow) and $N_{Cu}$
is the total number of Cu sites (here, $N_{Cu}=20\times 20=400$).
For $x<0.187$, there is enough space to avoid plaquette overlap
leading to $N_{4M}=N_{M}$, as seen in the $x=0.15$ figure.
The Methods section describes how the plaquettes are doped in figure 2.

Prior to describing our detailed model for
the anomalous normal state transport, we
summarize the cuprate doping phase diagram.
All cuprate superconductors have CuO$_2$ planes. When undoped, they
are insulating Heisenberg spin-1/2 antiferromagnets (AF)
with a N\'eel temperature $T_N\sim 200$ K
and spin-spin coupling $J_{dd}\approx\ 0.13$ eV \cite{Sulewski1990}.
The spin is localized on
planar Cu atoms in a d$^9$ configuration with the single unpaired
spin in the Cu \dxxyy\ orbital, where
the $x$ and $y$ axes point along the Cu$-$O bonds.
Doping of more than $x\approx 0.05$ holes per planar Cu
destroys the bulk N\'eel AF phase and
superconductivity occurs for $T<T_c$ where \tc\ is the superconducting
transition temperature. Above \tc, the material is metallic.
When $x$ is increased past $\approx 0.27$, 
superconductivity vanishes ($T_c=0$).  The highest
\tc\ occurs for $x\approx 0.16$ \cite{Tallon1995}.
The band structure consists of approximately
$2D$ CuO$_2$ bands with
Cu \dxxyy\ and O \psigma\ ($p_x$ and $p_y$)
orbital characters that cross the Fermi level \cite{Pickett1989}. The Fermi
surface is hole-like and centered around $k=(\pi/a,\pi/a)$ where
$a$ is the CuO$_2$ unit cell distance ($a\approx 3.8\ \mbox{\AA}$).

In our Plaquette model \cite{Tahir-Kheli2011},
doping creates holes in out-of-the-CuO$_2$-plane orbitals that
are localized in square
$4$-Cu-site plaquettes centered at the dopant.
This orbital character was shown by high quality ab-initio quantum mechanics
calculations on \lsco\ supercells \cite{Perry2002}.
These hole orbitals induce delocalization of the $4$ Cu \dxxyy\ and
$8$ O \psigma\ orbitals inside the doped plaquette. Each plaquette
can be thought of as a tiny piece of metal (see Supplement).

Figure 3 shows a Sr doped 4-Cu-site plaquette (as found in \lsco).
In figures 3a and 3b
the two degenerate out-of-plane states are shown with their respective
Jahn-Teller atomic distortions (blue arrows for planar O atoms and
black arrows for apical O atoms).  In the cuprates, no long-range static
distortion compatible with figure 3a or 3b is found.
Thus the coupling between these out-of-plane hole states
is strong, leading to a vibronic or
dynamical Jahn-Teller state being formed instead.

Figure 3c shows an instantaneous configuration of the out-of-plane
holes in a single CuO$_2$ plane.  The out-of-plane hole states
are shown by the red ``dumbbells".  Coupling between adjacent
plaquettes leads to correlation between the orientations
of neighboring dumbbells.
When plaquettes overlap as shown by the green squares in figure 3c,
the orbital degeneracy of the out-of-plane
holes shown in figures 3a and 3b is removed.  Thus overlapping plaquettes
have no dynamical Jahn-Teller state.

In figure 3d, the correlation energy
between neighboring dynamical
Jahn-Teller states is shown for plaquettes inside
a single CuO$_2$ plane, $\epsilon_{\mathrm{planar}}$,
and for plaquettes
in adjacent CuO$_2$ planes, $\epsilon_{\mathrm{perp}}$.  In the Supplement,
$\epsilon_{\mathrm{perp}}$ is estimated to be
$\sim 3.6\times 10^{-5}$ eV $\sim 0.42$ K
and $\epsilon_{\mathrm{planar}}$ is on the order 0.01 to 0.1 eV
($\sim 100-1000$ K).

For temperatures greater than $\sim 1$ K, there is no correlation
between dynamic Jahn-Teller states between planes.  In analogy
to the ``dynamical detuning" proposed for electron transport
normal to the CuO$_2$ planes \cite{Leggett1992,Turlakov2001},
the lack of correlation of dynamical Jahn-Teller states normal
to the plane disrupts the phase coherence of phonons
between CuO$_2$ layers.  Hence, the
phonon modes inside non-overlapping plaquettes
become strictly $2D$ for temperatures greater than
$\epsilon_{\mathrm{perp}}$.

Since momentum is not a good quantum number for
the $2D$ phonon modes inside the region of the non-overlapping plaquettes,
these modes have the same character as those in amorphous metals.
We show later that these amorphous $2D$ phonons lead to the low-temperature
linear resistivity (the $\alpha_1(0)$ term in figure 1).
Also, $\alpha_1(0)\propto N_{4M}$
because the number of amorphous $2D$ phonon modes
is proportional to the total number of Cu sites inside the
non-overlapping $4$-Cu-site plaquettes, $N_{4M}$.

The phonon modes occurring ``outside'' the non-overlapping plaquettes
are amorphous and $3D$.  The number of these modes is proportional to
$N_{Cu}-N_{4M}$.  These modes lead to the $T^2$ resistivity term,
as we show later.

A percolating pathway of plaquettes leads to
a delocalized metallic Cu \dxxyy\ and O \psigma\
band inside the percolating swath (yellow region in figure 2).
The undoped regions (not part of the doped metallic swath) remain
localized $d^9$ Cu spins with AF coupling.
Thus cuprates are intrinsically
inhomogeneous with a percolating metallic region and insulating AF regions.

Our proposed inhomogeneity appears prima facie to be at odds with
the recent observation of quantum oscillations in heavily overdoped
Tl-2201 cuprates \cite{Rourke2010,Vignolle2008,Bangura2010}.
In the Supplement, we estimate the Dingle temperature
arising from the inhomogeneity in our model and show that it
is compatible with these measurements.

The phonon modes that lead to superconductivity involve the planar
displacement of the
O atoms at the interface between the metal and AF regions (figures 4 and 5).
These longitudinal-optical (LO) O phonon modes
are the softened modes seen by neutron scattering
\cite{Pintschovius2006,Reznik2006,Pintschovius2005}
for the superconducting range of dopings
for momenta along the $(\pi,0)$ and $(0,\pi)$ directions.
They are different from the Jahn-Teller modes (figure 3a,b) that lead to the
linear $T$ term in $\rho$.
Figure 6 shows how these modes lead to the observed
d-wave (\dxxyy) superconducting gap.

To complete the derivation of the results in figure 1, we must
show that phonon modes inside the non-overlapping plaquettes produce
the anisotropic $\cos^2(2\varphi)T$ scattering rate and that the
remaining phonons contribute the $T^2$ isotropic term.
In a typical metal with full translational symmetry, the
scattering rate for a metallic electron with momentum $k$ is
\cite{Ziman1972}
$1/\tau(k)\sim \sum_{k',q}(\hbar/M\omega_q) q^2
(1-\cos\theta)n_B(\omega_q)\delta(k'-k-q)$
where the sum is over $k'$ on the Fermi surface and $q$ is the phonon momentum.
$M$ is the nuclear mass, $\omega_q$ is the phonon energy,
$\theta$ is the angle between $k$ and $k'$, and
$n_{\mathrm B}(\omega)=1/(e^{\hbar\omega/k_B T}-1)$
is the Bose-Einstein distribution
($k_B$ is Boltzmann's constant).
The delta function maintains momentum conservation.  
The $(1-\cos\theta)$ term becomes $T$ dependent at low $T$ because
$n_B(\omega)$ restricts the scattering to small-angles ($\theta\ll 1$).

All phonons can scatter $k$ to $k'$ in cuprates
because of their intrinsic inhomogeneity.
Hence, the $(1-\cos\theta)$ term
in $1/\tau(k)$
has no $T$ dependence.  In this case \cite{Bergmann1971},
$1/\tau(k)\sim \sum_{k',\lambda}(\hbar/M\omega_\lambda)
(k'-k)^2 n_{\mathrm B}(\omega_\lambda)$
where $\lambda$ is summed over all phonon modes and the delta function
has disappeared.
The number of excited phonon modes at temperature
$T<\Theta_{\mathrm{Debye}}$ is
$\sim T^d$ where $d$ is the dimension of the phonon modes \cite{Cote1981}.
At high $T$, all phonon modes are excited.
At low temperatures, $1/\tau\sim (1/\omega)T^d\sim T^{d-1}$, because
$\hbar\omega\sim k_B T$.
Our expression for $1/\tau$ is identical to the Debye-Waller factor
\cite{Ziman1972,Cote1981} that appears in the resistivity of
amorphous metals (see Supplement for details).
At high $T$, $1/\tau\sim T$ because
all phonon modes become thermally accessible
and the occupation number for each mode, $n_{\mathrm B}$,
is proportional to $T$.

The $2D$ Jahn-Teller modes in figure 3 have $d=2$, leading to
$1/\tau\sim T$. This linear scattering rate persists for
$T>\epsilon_{\mathrm{perp}}\sim 0.4$ K.
The remaining $3D$ modes lead to $1/\tau\sim T^2$.
Since the phonon modes are amorphous, $1/\tau(k)$ is
independent of $k$ for the $3D$ $T^2$ modes.
The $2D$ Jahn-Teller
modes are derived from the vibronic distortions in figure 3.
These modes raise the Cu orbital energy for two diagonal Cu
atoms in a plaquette and lower the orbital energy for the other
two Cu atoms.  Hence, their scattering is modulated by
an envelope function $\approx\cos^2(2\varphi)$.
Thus $1/\tau(k)\sim \cos^2(2\varphi)T$.
This modulation of $1/\tau(k)$ does not change the total scattering around the
Fermi surface.  Instead, it merely redistributes its weight
around the surface.  Hence, the proportionality
constant in front of the expressions for
$\alpha_1(0)$ and $\alpha_1(\infty)$ is the same.

Thus the doping and temperature dependence of the resistivity,
Hall effect, and magnetoresistance of cuprates is due to
the intrinsic inhomogeneity of these materials.  This inhomogeneity
arises from the dopant atoms and leads to a
percolating metallic region and insulating antiferromagnetic
regions.
The percolating metallic region is comprised of 4-Cu-site plaquettes
that are centered around dopants.  The plaquettes may
overlap each other.  We show that the ratio
of non-overlapping plaquettes to the total number of Cu atoms
equals the ratio of the low and high temperature
linear $T$ terms in the resistivity, $\alpha_1(0)/\alpha_1(\infty)$, with
no adjustable parameters.  We calculate this ratio and find very good
agreement with experiment.
The resistivity is shown to be due to
phonon scattering.  The special ``crossover" doping, $x\approx 0.19$, for
$\alpha_1(0)$ and $\alpha_1(\infty)$
occurs at the doping when plaquettes first begin to overlap.
We calculate this value to be $x=0.187$ with no adjustable parameters.
The low doping value of $\alpha_1(0)$ and the high doping value
of $\alpha_1(\infty)$ are found to be exactly the same in our
model.  This is observed by experiments within the error bars.
We find an excellent fit to the doping evolution of
$\alpha_1(0)$ and $\alpha_1(\infty)$
with one adjustable multiplicative constant.
We also provide a physical picture for the origin of the observed softened
LO O phonon mode and show that it leads to a d-wave superconducting gap.

We conclude that the anomalous normal state transport
in cuprates provides strong
evidence for intrinsic inhomogeneity, metallic percolation,
phonon superconducting pairing, and phonon normal state scattering.

\section*{Methods}
We dope plaquettes using a simple model for the
Coulomb repulsion of the dopants.  Since the dopants reside in
the metallic region, the repulsion is screened over a
short distance.  Thus we assume plaquettes are randomly
doped without overlap.  This doping is possible up to $x=0.187$.
For $x>0.187$, it is impossible to add dopants without overlapping
another plaquette.  Above $x=0.187$, the
most energetically favorable location for a dopant is on a site
that dopes three undoped d$^9$ Cu atoms and one previously
doped Cu atom.
At $x=0.226$, it is no longer possible to locate three undoped
d$^9$ Cu atoms that are part of a 4-site square. Hence,
two undoped d$^9$ Cu's and two previously doped Cu's are doped up to $x=0.271$.
At this point, there are no adjacent pairs of d$^9$ spins.
Single undoped d$^9$ Cu doping occurs up to $x=0.317$.
Above this doping, there are no remaining undoped d$^9$ Cu's.
The green squares in figure 2 show all the
plaquettes that overlap another plaquette (share at least one Cu atom).
The percolating metallic region has become the whole
crystal.  We have not computed dopings beyond this value ($x=0.317$).
For the results in figure 1, we doped a
$10^3 \times 10^3$ square CuO$_2$ lattice $10^3$ different times.
$N_{4M}/N_{Cu}$, and $N_{M}/N_{Cu}$ values were obtained by averaging
over the ensemble (see figures S1 and S2).

\vspace{.1in}

\noindent \textbf{Acknowledgements.} The author is grateful to Andres
Jaramillo-Botero and Carver A. Mead for many stimulating discussions.

\pagebreak

\begin{figure}[tbp]
\centering \includegraphics[width=1.0\linewidth]{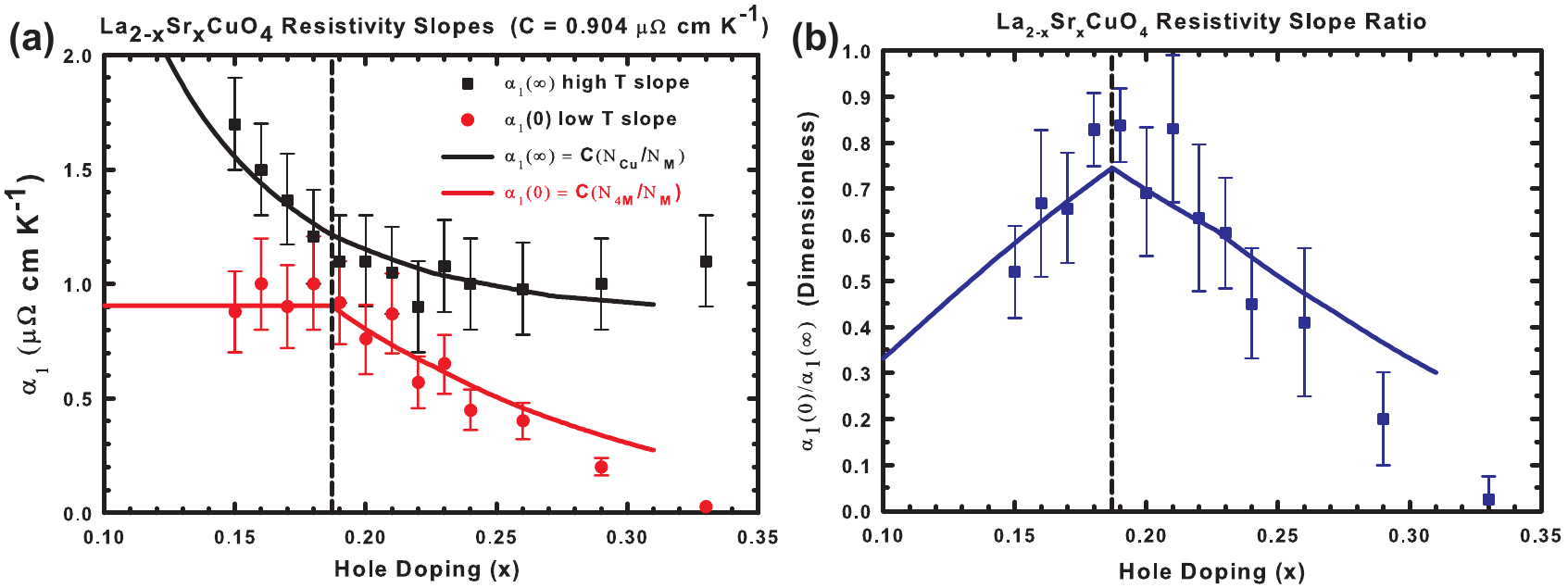}
\caption{Doping evolution of the linear $T$ coefficients of the resisitivity
at high and low temperatures
($\alpha_1(0)$ and $\alpha_1(\infty)$).
(a) Experiment \cite{Hussey2011a} versus theory.
$\alpha_1(0)=C(N_{4M}/N_{M})$ and
$\alpha_1(\infty)=C(N_{Cu}/N_{M})$ where
$N_{M}$ is the total number of metallic Cu sites (yellow in 
figure 2), $N_{4M}$ is the number of Cu atoms inside the
non-overlapping 4-Cu-site plaquettes (black squares with
the number ``4'' in figure 2), and
$N_{Cu}$ is the total number of planar Cu sites
(see Figures S1 and S2).
The low and high temperature linear $T$ scattering rates
arise from phonons and are given by $1/\tau(0)\sim N_{4M}T$ and
$1/\tau(\infty)\sim N_{Cu}T$, respectively.
Dividing by the total number of
charge carriers, $N_{M}$, leads to $\alpha_1(0)$ and $\alpha_1(\infty)$.
For $C=0.904\ \mu\Omega$-cm/K,
the RMS and maximum absolute errors are
$0.017$ and $0.17\ \mu\Omega$-cm$/$K, respectively.
(b), plot of the ratio
$\alpha_1(0)/\alpha_1(\infty)=N_{4M}/N_{Cu}$.
$C$ cancels out here, so there are zero adjustable parameters.
For doping $x<0.187$, $N_{4M}=N_{M}$ because there are no
overlapping plaquettes.  Hence, $\alpha_1(0)=C$.
The start of plaquette overlap at $x=0.187$ (vertical dashed line)
leads to the sharp discontinuity in $\alpha_1(0)$ there.
Since $N_{M}\rightarrow N_{Cu}$
with increasing doping, $\alpha_1(\infty)$ at high doping tends to
the same constant $C$, as observed.
(a) and (b) are the main results of the paper.
}
\end{figure}

\pagebreak

\begin{figure}[tbp]
\centering \includegraphics[width=0.60\linewidth]{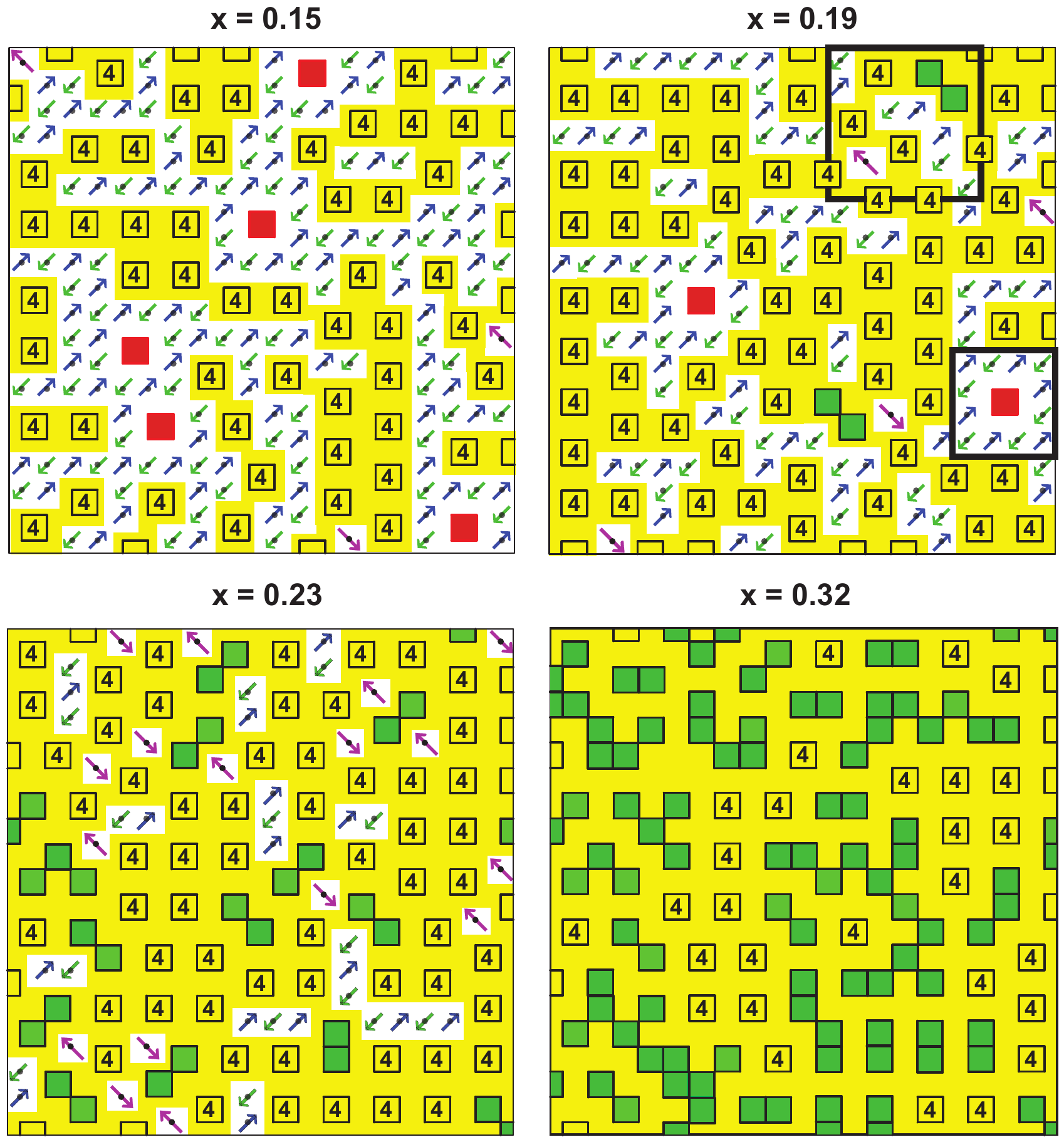}
\caption{The distribution of plaquettes (non-overlapping and overlapping)
on a $20\times 20$ lattice in a single CuO$_2$ plane.
Each black, green, pink, blue, and red square represents a plaquette
centered at a dopant (Sr in \lsco).
The corners of the squares are planar Cu sites. The O atoms are not shown.
A Cu \dxxyy$/$O \psigma\ 
metallic band forms inside the percolating region of the
plaquettes and is shown in yellow (see Supplement).
The non-overlapping plaquettes (black squares with the number ``$4$''
inside) will have dynamic Jahn-Teller distortions as shown in figure
3.  The number of these amorphous $2D$ phonon modes is proportional to
the number of Cu sites inside these plaquettes, $N_{4M}$.
$N_{M}$ is the total number of metallic Cu sites (yellow).
$N_{Cu}$ is the total Cu sites. Here, $N_{Cu}=20\times 20=400$.
The arrows are undoped d$^9$ Cu spins that are
antiferromagnetically (AF) coupled.
Plaquettes do not overlap for $x<0.187$, leading to $N_{4M}=N_{M}$
(see $x=0.15$ figure).
For $x>0.187$, overlap is unavoidable (green squares). See Methods section.
The red squares are isolated plaquettes with no plaquette neighbors.
These plaquettes lead to the pseudogap (PG) \cite{Tahir-Kheli2011}.
They are not connected to the percolating ``metallic'' swath.
The $6\times 6$ black square in the upper right of $x=0.19$ is expanded
in figure 3c to show the $2D$ Jahn-Teller phonon modes and in figure 4
to show the O atom distortions.  The $4\times 4$ square in the lower
right is expanded in the Supplement to show the vicinity of an isolated
plaquette (red square).
}

\pagebreak

\end{figure}
\begin{figure}[tbp]
\centering \includegraphics[width=0.55\linewidth]{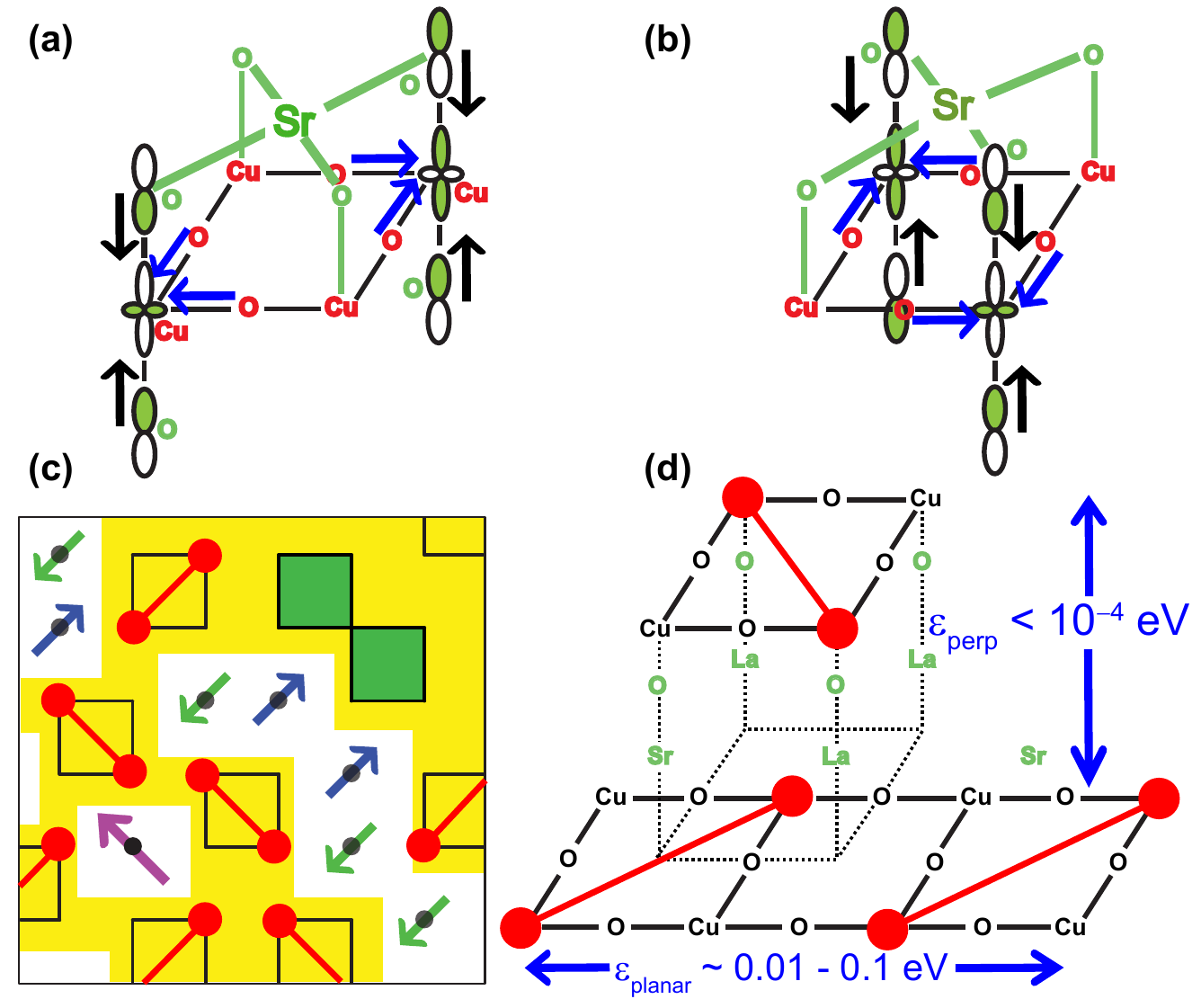}
\caption{Dynamic Jahn-Teller character of doped plaquettes and their
atomic distortions.
(a) and (b), the localized out-of-the-CuO$_2$-plane hole character
surrounding a Sr dopant in \lsco.  There are two degenerate states.
The blue and black arrows show local distortions that stabilize these
orbital states.
A resonance of these two configurations (a vibronic or dynamic
Jahn-Teller effect) occurs.
These hole orbitals have Cu
\dzz\ and apical O \pz\ character (vertical white-green orbitals)
that induce the planar Cu \dxxyy\ and O \psigma\ 
orbitals inside this plaquette (shown in yellow in (d)) to
delocalize over the $4$-Cu-sites inside the plaquette.
When the plaquettes percolate through the crystal (yellow region
in (c) and figure 2), a planar Cu/O \dxxyy$/$\psigma\ metallic band is formed
inside the yellow region \cite{Perry2002,Tahir-Kheli2011} (see Supplement).
(c) Looking down at a $6\times 6$ square in the CuO$_2$ plane taken
from the upper right corner of $x=0.19$ doping in figure 2.  The red
``dumbbells'' show the instantaneous orientation of the out-of-plane
orbitals in the square.
The degeneracy in (a) and (b) is split in the two overlapping
plaquettes in the upper right (green squares).
No dynamic Jahn-Teller states are formed there.
(d) The coupling energy between neighboring ``dumbbells" inside
a CuO$_2$ plane and between neighboring planes.  The coupling energy
between planes is very small,
$\epsilon_{\mathrm{perp}}\sim 3.6\times 10^{-5}$ eV $\sim 0.42$ K
(see Supplement for the calculation).
For $T>\epsilon_{\mathrm{perp}}\sim 0.42$ K,
there is no correlation of the dynamic Jahn-Teller states in
adjacent CuO$_2$ planes.  Hence, the phonon modes inside non-overlapping
plaquettes lose phase coherence between the planes and become
strictly $2D$ \cite{Leggett1992,Turlakov2001}.  Within each CuO$_2$ plane,
these modes are amorphous (momentum is not a good quantum number).
Amorphous $2D$ phonons
lead to the low-temperature linear $T$ coefficient, $\alpha_1(0)$,
as shown in the text.
}
\end{figure}

\pagebreak

\begin{figure}[tbp]
\centering \includegraphics[width=0.8\linewidth]{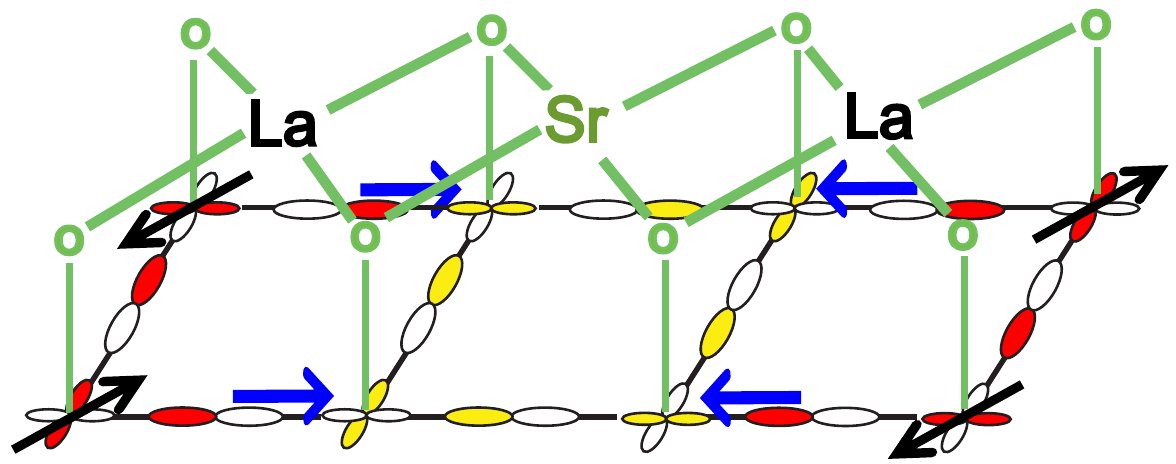}
\caption{
3D picture of the distortion of the O atoms between the
metallic (yellow) and AF (red) regions.
These modes are shown in greater detail in figure 5 and
lead to d-wave superconductivity as shown in figure 6.
}
\end{figure}

\pagebreak

\begin{figure}[tbp]
\centering \includegraphics[width=0.6\linewidth]{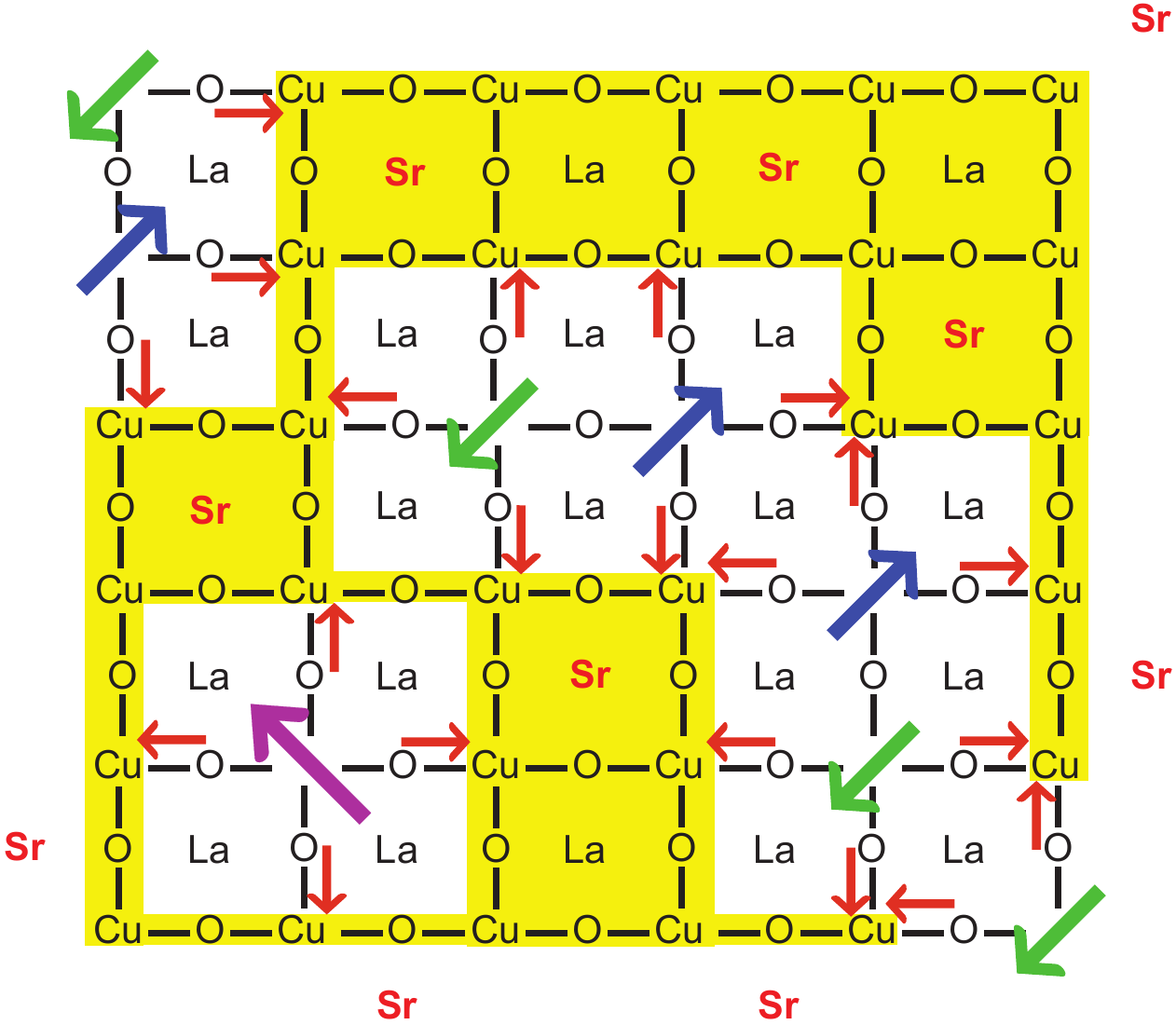}
\caption{The AF/Metal interfacial O phonon modes.
Fully expanded $6\times 6$
black square in the upper right corner of $x=0.19$ doping from figure 2.
The interfacial O phonon modes (red arrows)
are the origin of the observed softened phonon
longitudinal optical (LO) oxygen modes seen by neutron scattering
with wavevectors along the Cu-O bond directions $[(0,0)$ to $(\pi,0)$ and
$(0,\pi)]$ for the superconducting range of dopings
\cite{Pintschovius2006,Reznik2006,Pintschovius2005}.
These modes lead to a d-wave superconducting gap as shown in figure 6.
}
\end{figure}

\pagebreak

\begin{figure}[tbp]
\centering \includegraphics[width=0.8\linewidth]{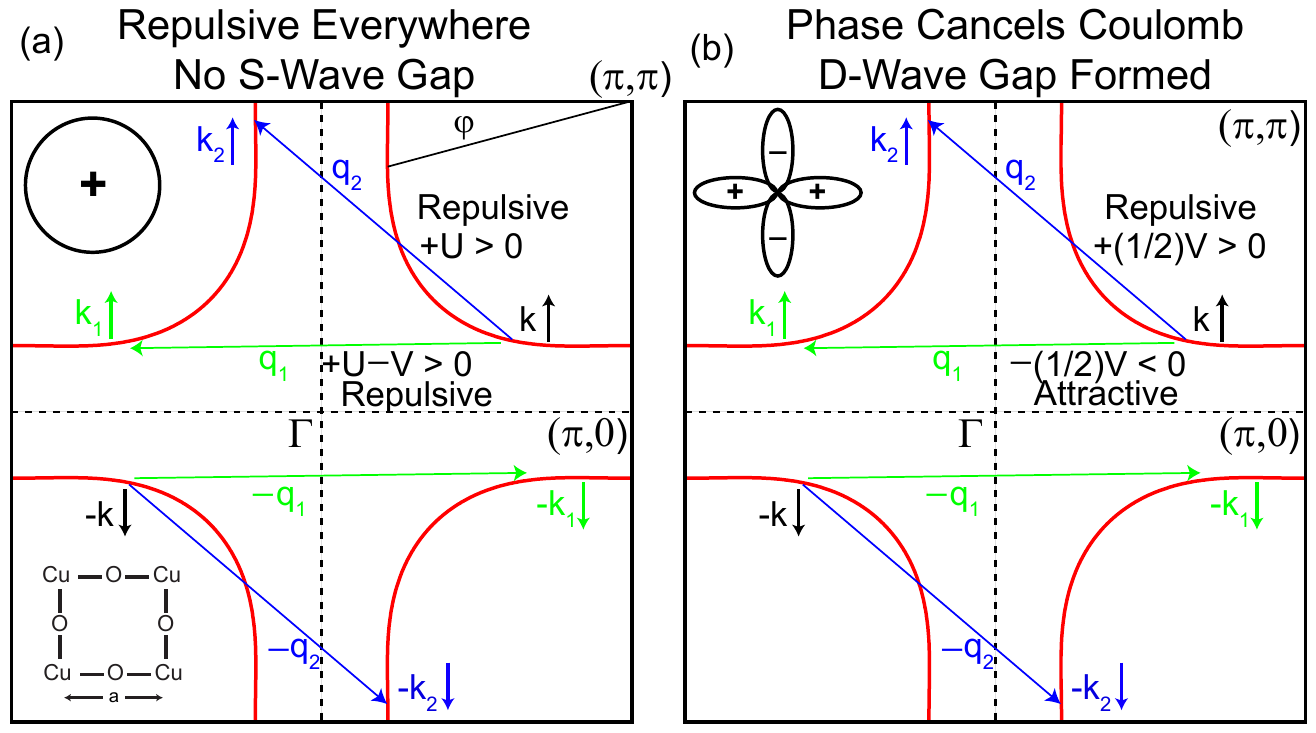}
\caption{
Schematic diagram of the cuprate Fermi surface showing how a
strong attractive phonon pairing along the Cu$-$O bond directions leads to the
observed d-wave (\dxxyy) superconducting gap.  (a) and (b) show the
hole-like Fermi surface centered at $(\pi/a,\pi/a)$
where $a\approx 3.8\ \mbox{\AA}$ is the unit cell distance
(the $\Gamma$-point is occupied).
The Cu$-$O unit cell in the lower left corner of (a) shows the
orientation of the Brillouin zone relative to the crystal lattice.
In both (a) and (b), a $(k\up,-k\down)$ Cooper pair in black is scattered by
a phonon with momentum
$q_1$ or $q_2$ to $(k_1\up,-k_1\down)$ in green and
$(k_2\up,-k_2\down)$ in blue, respectively.
A Coulomb repulsion, ($+U>0$), occurs for both $q_1$ and $q_2$.
$q_1$ has
an additional phonon-mediated attraction, $–V<0$, from the AF/metal
interfacial O atoms in figures 4 and 5.
For an isotropic s-wave gap, there is a net repulsion everywhere
(upper left corner of (a)).  Superconductivity cannot occur for this gap
symmetry.  For a \dxxyy\ gap (upper left of (b)),
the average coupling around the Fermi surface of $+U-(1/2)V$ integrates
to zero leaving a net attractive $-(1/2)V$ pairing along the Cu$-$O bond
directions and a repulsive $+(1/2V)$ coupling for
$q\approx(\pi/a,\pi/a)$.
A \dxxyy\ gap can lower the energy.  Thus the modes in figures 4 and 5 lead to
the observed d-wave gap.  The idea of a d-wave gap
arising from the softened oxygen LO phonon modes along the Cu$-$O bond
directions has been suggested previously
\cite{Phillips2003}.
In particular, Phillips
\cite{Phillips1989,Phillips2007}
argued for softened modes occuring at the interface
of the AF and metallic regions.
What is new here, is that the physical origin of these phonon modes
is provided along with its doping evolution.
}
\end{figure}

\clearpage

\newpage

\setcounter{figure}{0}
\setcounter{page}{1}

\vspace*{5cm}

\begin{center}
{\bf \large Supplementary Information for\\``Resistance
\\ of\\ High-Temperature Cuprate Superconductors''}

{\rm Jamil Tahir-Kheli}


{\it Department of Chemistry}

{\it Beckman Institute (MC 139-74)}

{\it California Institute of Technology}

{\it Pasadena, CA 91125}
\end{center}

\newpage

\baselineskip24pt

\subsection*{Metal Plaquette Counting as a Function of Doping}
Figure S1 shows the calculated number of Cu atoms inside the
metallic region, $N_{M}$, and the number of metallic Cu atoms
that reside in non-overlapping plaquettes, $N_{4M}$. In figure 2
of the main text, the number of Cu sites in the
black squares with ``4'' inside is
$N_{4M}$ and the yellow region of the same figure is $N_{M}$.
The doping evolution of these two numbers determines the
low and high $T$ linear coefficients of the resistivity shown in figure 1
($\alpha_1(0)$ and $\alpha_1(\infty)$).

Figure S2 shows the doping evolution of the two different types of
non-overlapping plaquettes.  Only the non-overlapping plaquettes
inside the metallic regions (black squares with ``4'' inside
in figure 2 of the main text) contribute to the low $T$ coefficient
of the resistivity ($\alpha_1(0)$).
There are additional non-overlapping plaquettes
in figure 2 outside of the metallic region.  They are shown by red
squares in the figure and are isolated plaquettes.  The number
and spatial distribution of these isolated plaquettes determines
the pseudogap (PG) and its doping evolution
as shown in reference \cite{Tahir-Kheli2011}.

\subsection*{Cu/O $\mathbf{d_{x^2-y^2}/p_\sigma}$\ 
Band Delocalization Inside the Metallic Region}

Figure S3 shows the shift in the orbital energy of
the metallic Cu \dxxyy\ orbital and the neighboring O \psigma\ orbital
due to the dopant induced localized out-of-plane hole.
Since the \dxxyy\ orbital energy is lowered more than \psigma, delocalization
can occur, leading to a Cu/O \dxxyy/\psigma\ band inside the
percolating metallic region. Further details are provided in
reference \cite{Tahir-Kheli2011} and its supplementary
information.

\subsection*{Ab-Initio DFT Calculations on
La$\mathbf{_{2-x}}$Sr$\mathbf{_x}$CuO$\mathbf{_4}$ Supercells}

Pure density functionals such as local density (LDA)
[S1, S2, 17]
and gradient-corrected functionals (GGA, PBE, etc)
[S3]
obtain a metallic ground
state for the undoped cuprates rather than an antiferromagnetic insulator
because they underestimate the band gap due
to a derivative discontinuity of the energy with respect to the number of
electrons
[S4, S5].
In essence, the LDA and PBE functionals include too much
self-Coulomb repulsion.  This repulsion leads to more delocalized electronic
states.  Removing this extra repulsion is necessary in order
to obtain the correct localized antiferromagnetic
spin states of the undoped cuprates.

The self-repulsion problem with LDA has been known for a long time
[S6].
Very soon after the failure of LDA to obtain the undoped insulating
antiferromagnet for cuprates, several approaches were applied to correct
the flaws in LDA for La$_2$CuO$_4$.
Using a self-interaction-corrected method
(SIC-LDA) [S6], Svane [S7] achieved
spin localization with an indirect band gap of 1.04 eV, and Temmerman,
Szotek, and Winter [S8]
found a band gap of 2.1 eV.  Using an LDA + U method,
Czyzyk and Sawatzky [S9]
obtained 1.65 eV.   In all of these calculations
on the undoped cuprates, an increase in out-of-plane orbital character
was noted in states just below the top of the valence band.  Calculations
with explicit dopants such as Sr in La$_{2-x}$Sr$_x$CuO$_4$ were not done.

Our ab-initio calculations [S10, 18]
were performed using the hybrid density functional, B3LYP.
Due to its remarkable success on molecular systems,
B3LYP has been the workhorse density functional
for molecular chemistry computations for almost 20 years
[S11, S12].
For example [S11],
B3LYP has a mean
absolute deviation (MAD) of 0.13 eV, LDA MAD = 3.94 eV, and PBE MAD = 0.74 eV
for the heats of formation, $\Delta H_f$, of the 148 molecules in the
extended G2 set [S13, S14].
B3LYP has also been found to predict excellent band gaps
for carbon nanotubes and binary and ternary semiconductors relevant to
photovoltaics and thermoelectrics
[S15, S16].

The essential difference between B3LYP and all pure density
functionals is that 20\% exact Hartree-Fock (HF) exchange is included.
B3LYP is called a hybrid functional because it includes exact HF exchange.
The HF exchange
removes some of the self-Coulomb repulsion of an electron with itself
found in pure DFT functionals.  A modern viewpoint of the reason for the
success of hybrid functionals is that inclusion of some exact
Hartree-Fock exchange compensates the error for fractional charges that
occur in LDA, PBE, and other pure density functionals
[S17].
The downside to
using hybrid functionals is they are computationally more expensive than
pure density functionals.

Our B3LYP calculations reproduced the experimental $2.0$ eV band gap for
undoped La$_2$CuO$_4$ and also had very good agreement for the
antiferromagnetic spin-spin coupling, $J_{dd}=0.18$ eV (experiment is
$\approx 0.13$ eV) [S10].
We also found substantial out-of-plane apical O $p_z$ and Cu
$d_{z^2}$ character near the top of the valence band in agreement with
LDA + U and SIC-LDA calculations.

We also performed B3LYP calculations on La$_{2-x}$Sr$_x$CuO$_4$ for
$x = 0.125$, $0.25$, and $0.50$ with explicit Sr atoms using large
supercells [18].
Regardless of the doping value, we always found that the
Sr dopant induces a localized hole in an out-of-the-plane orbital that
is delocalized over the four-site region surrounding the Sr as shown in
figures 3 and S3.
This result is in contrast to removing an electron from the
planar Cu \dxxyy/O \psigma\ as predicted by LDA and PBE.

Our calculations found that the apical O's in the doped CuO$_6$
octahedron are asymmetric anti-Jahn-Teller distorted.   In particular,
the O atom between the Cu and Sr is displaced $0.24\ \mbox{\AA}$
while the O atom between the Cu and La is displaced $0.10\ \mbox{\AA}$.
XAFS measurements [S18]
find the apical O displacement in the vicinity of a Sr to
be $\approx 0.2\ \mbox{\AA}$.

\subsection*{Compatibility of Our Proposed Inhomogeneity and the
Quantum Oscillations in Overdoped Tl-2201}

De Haas-van Alphen quantum oscillations have recently been observed
in heavily overdoped Tl-2201
\cite{Rourke2010,Vignolle2008,Bangura2010}.  These authors concluded
that there in no microscopic inhomogeneity in the CuO$_2$ planes
over a minimum of $1200\ \mbox{\AA}$.  Hence, these experiments appear
to contradict our model of microscopic inhomogeneity in the CuO$_2$
planes.

Here, we estimate the amount of scattering to be expected from the inhomogeneity in our model and show it is compatible with the data in Rourke et al
\cite{Rourke2010}.
Rourke et al find quantum oscillations in
3 of the approximately 100 samples tested.
Two of samples showing oscillations had doping
at $\approx 0.31$ doping and the remaing one had doping at $\approx 0.26$.
Significantly, no samples with lower doping showed quantum
oscillations.

Using the
doping methodology described in the Methods section of our paper, we find that
4.5\% of the planar Cu atoms are localized d$^9$ spins at 0.26 doping and
0.6\% of the planar Cu atoms are localized d$^9$ spins at 0.31 doping.
At these two dopings, only isolated d$^9$ spins remain in our model.
Below $\approx 0.26$ doping, some of the d$^9$ Cu spins neighbor
other d$^9$ Cu spins.  These d$^9$ Cu atoms are no longer isolated.

In his classic book, Shoenberg [S19]
states that Dingle
temperatures much above 5K smear out quantum oscillations so that they are not
resolvable (page 62). Shoenberg also states (page 62 and Table 8.1) that
every 1\% impurity typically adds $10-100$ K to the Dingle temperature.
Thus it is possible the $\approx 0.6$\% ``impurities" we expect for the
0.31 doped samples may still lead to observable quantum oscillations, but
the 4.5\% impurities at 0.26 doping appears to be too large to be
compatible with experiment.

The flaw in the above simplified analysis is that a localized Cu
d$^9$ ``impurity" is not a typical impurity. The charge distribution around
a metallic and localized d$^9$ Cu are virtually identical in the
CuO$_2$ planes in cuprates.
For both Cu atoms, the Cu has approximately $+2$ charge with a hole in the
$d_{x^2-y^2}$ orbital.
The d$^9$ Cu impurity is not a charged impurity.  Instead, it is a
neutral impurity with exactly the same sized valence orbitals as the
metallic Cu atom.
Since the impurity scattering cross section (and hence the scattering length)
scales as the square of the difference in the potential between these two
types of Cu atoms, the scattering can be up to two orders of magnitude
smaller than a typical impurity, as we show below.  Thus 4.5\% d$^9$
Cu impurities may lead to the same amount of scattering as 0.045\%
atomic substitutions of different atoms.  At $10-100$ K Dingle temperature
per 1\% impurity, this leads to a Dingle temperature of $0.45-4.5$ K
for the 0.26 doped sample.

We can use the Friedel sum rule \cite{Ziman1972} to quantify
this physical argument.  The Friedel sum rule incorporates the
response of the metallic electrons around an impurity to screen out the
impurity charge distribution and maintain a constant Fermi energy throughout
the crystal. The relevant Cu orbitals near the Fermi level are
$d_{x^2-y^2}$, $d_{z^2}$, and $4s$ with symmetries $B_{1g}$, $A_{1g}$,
and $A_{1g}$, respectively.  The charge difference between the impurity Cu
atom and the metallic Cu atom is $Z=0$, leading to
$Z=0=\frac{2}{\pi}[\delta_A + \delta_B]$
where $\delta_B$ is the $d_{x^2-y^2}$ phase shift at the Fermi level
and $\delta_A$ is the combined $d_{z^2}$ and $4s$ phase shifts at the
Fermi level.  Since there is predominantly
$d_{x^2-y^2}$ character at the Fermi level,
the $d_{z^2}$ and $4s$ phase shifts should be small, $\delta_A\sim 0.1$.

The Friedel sum rule therefore forces the $d_{x^2-y^2}$ phase shift to be
small, $\delta_B\sim 0.1$ rather than a $\delta_B\sim 1$ for a ``standard"
impurity arising from atomic substitution.

Since the scattering cross section scales as
$\sin^2\delta_B$ \cite{Ziman1972}, the scattering cross section may be two
orders of magnitude smaller than a typical impurity.
Thus we have shown that the scattering length expected from the
inhomogeneity of our plaquette model leads to Dingle temperatures from
$0.45-4.5$ K at 0.26 doping and $0.006-0.6$ K at 0.31 doping.

Rourke et al measure the attenuation of the oscillations due to the Dingle
temperature, but they do not quote their Dingle temperature.  We can extract
their Dingle temperature in two ways.

First, the Dingle temperature, $x_D$, is given by

\begin{equation}
k_B x_D=\frac{\hbar}{2\pi\tau}
\end{equation}

\noindent
where $k_B$ is Boltzmann's constant and $\tau$ is the scattering time.
Multiplying the top and bottom of the right-hand-side of this equation by
the Fermi velocity, $v_F$, we obtain

\begin{equation}
k_B x_D=\frac{\hbar v_F}{2\pi l_0}
\end{equation}

\noindent
where $l_0$ is the scattering length.  $l_0=400\ \mbox{\AA}$ in Rourke et al.
$v_F\approx 1-3\times 10^7$ cm/s in cuprates \cite{Hussey2003}.  For
$v_F=2\times 10^7$ cm/s, $x_D=6.1$ K.

Second, we may use

\begin{equation}
k_B x_D=\left(\frac{m_{elec}}{m_{therm}}\right)
\left(\frac{2\mu_B}{2\pi}\right)
\left(\frac{2\hbar F_0}{e}\right)^{1/2}
\left(\frac{1}{l_0}\right)
\end{equation}

\noindent
where $m_{therm}/m_{elec}$ is the ratio of the band electron mass measured
by the temperature dependence of the oscillation amplitude ($\sim 5$ in
Rourke et al), $\mu_B$ is the Bohr magneton, $F_0$ is the oscillation
frequency ($1.8\times 10^4$ Tesla in Rourke et al).
This leads to $x_D=5.2$ K.

Therefore, estimates of the scattering from our plaquette induced
inhomogeneity lead to a Dingle temperature compatible with observations.

In our model, below $\approx 0.26$ doping, there are d$^9$ Cu atoms that are
no longer isolated.  These Cu atoms form antiferromagnetic
regions in the material.  We anticipate that the scattering from
these ``non-isolated" regions will lead to greater scattering than
isolated d$^9$ Cu atoms.  This may explain why no quantum oscillations
were found for dopings less than $\approx 0.26$.
 
Despite the calculations done here, further work needs to be done to
obtain precise numbers for the scattering cross section from an isolated
d$^9$ Cu spin.
What we have shown here is that it is most likely that the inhomogeneity
we propose in this paper does not contradict the quantum oscillations
found in heavily overdoped Tl-2201.

\subsection*{The Debye-Waller Factor and Amorphous/Glassy Metals}

For liquid metals, Ziman and Faber [S20, S21]
showed that
the $T$ dependence of the structure
factor, $S(K)$, evaluated at $K=2k_F$, where $k_F$ is the Fermi
momentum determines the $T$ dependence of the resistivity.
Their resistivity expression is given by [S20, 28],
\begin{equation}
\rho(T)=\frac{12\pi\Omega}{e^2\hbar v^2_F}
\int_0^1 d\left(\frac{K}{2k_F}\right)
\left(\frac{K}{2k_F}\right)^3
S(K,T)|t(K)|^2.
\end{equation}
Here $v_F$ is the Fermi velocity, $t(K)$ is the scattering
matrix element, and $\Omega$ is the volume.

The integral for $\rho$ is dominated by the maximum momentum
transfer across the Fermi surface, or $2k_F$.  Thus the $T$
dependence of the structure factor, $S(2k_F,T)$, determines
the temperature dependence of $\rho$.

In amorphous/glassy metals, the structure factor is dominated
by the ``zero-phonon'' elastic term that is given by
$S(K,T)=S(K)e^{-2W(K)}$ where $S(K)$ is the static structure
factor and $W(K)$ is the Debye-Waller factor arising from the
nuclear motion at temperature $T$ [S22, S23, S24, S25].
Here, $K=2k_F$.
Using $S(K,T)\approx S(K)[1-2W(K)]$, we conclude that the
temperature dependence of the $W(K)$ determines the temperature
dependence of $\rho(T)$.

The Debye-Waller factor is given by $W(K)=(1/2)\sum_q |K\cdot U_q|^2$
where the sum is over all phonon modes $q$ and $U_q$ is the
amplitude for the $q$ mode.  $U_q$ can be determined from
$(1/2)M\omega_q^2|U_q|^2\sim (1/2)n(\omega)\hbar\omega_q$ \cite{Ziman1972}
where $M$ is the nuclear mass, $\hbar\omega_q$ is the energy 
of the phonon mode $q$, and $n_q$ is the Bose-Einstein
occupancy of the phonon
mode, $n_q=1/(e^{\beta\hbar\omega_q}-1)$.
From this expression, we find the form of the the Debye-Waller factor to be,
\begin{equation}
W(K)\sim K^2 \int\frac{n_q}{\omega_q}d^3q.
\end{equation}
The expression for $W(K)$ has exactly the same form that we
obtained from the matrix element argument in the main text
\cite{Bergmann1971}.

The Bose-Einstein occupation, $n_q$,
cuts off the integral at $\sim k_B T$, where $k_B$ is
Boltzmann's constant.  Since $\omega_q=cq$, where $c$ is the speed
of sound, $W(K)\sim T^2$.  For $2D$ phonon modes, the integral
is over $d^2q$ instead of $d^3q$ leading to $W(K)\sim T$.

For $2D$, the Debye-Waller factor diverges as
$\int d\omega/\omega\sim \log T$
as $T\rightarrow 0$.  In fact, the $2D$ phonon modes
are not strictly $2D$ for $T<\epsilon_{\mathrm{perp}}\sim 0.4$ K.
Less than
$0.1$ K, these modes are $3D$ leading to a finite value for the
Debye-Waller factor.  For $T>0.1$ K, the Debye-Waller factor
is linear in $T$ leading to a linear resistivity.

\subsection*{Estimating the Coupling Energy of Dynamical Jahn-Teller
Plaquettes in Neighboring CuO$_2$ Planes}

The coupling energy, $\epsilon_{\mathrm{perp}}$, between two dynamical
Jahn-Teller plaquettes in adjacent planes is given by the Coulomb
repulsion energy difference between the different orientations
of the plaquette hole orbitals.  The details of the energy
difference expression are shown in Figure S4.

The Thomas-Fermi screening length, $\lambda$, is estimated from
the expression \cite{Ziman1972},

\begin{equation}
\frac{1}{\lambda^2}=4\pi e^2 N(0)
\end{equation}

\noindent
where $e$ is the electron charge and $N(0)$ is the density of
states.  There is approximately one metallic electron per unit
cell leading to a density of states of approximately one state per
eV per unit cell.  The volume of the unit cell is,
$\Omega_{cell}=a^2 d$, where $a=3.8\ \mbox{\AA}$ and $d=6.0\ \mbox{\AA}$.
Therefore, $N(0)\approx 87.0$ eV$^{-1} \mbox{\AA}^{-3}$ and
$\lambda=0.69\ \mbox{\AA}$.

Substituting these numbers into the expression shown in figure S4
leads to
$\epsilon_{\mathrm{perp}}=3.6\times 10^{-5}$ eV $=0.42$ K.

\subsection*{Oxygen Modes Around an Isolated (Red) Plaquette}

The red arrows in figure S5 show the softened phonon mode for the
O atoms between the AF spins and the delocalized electrons inside the
isolated plaquette.

\subsection*{References}

\noindent {\it S1.}
Yu, J. J., Freeman, A. J., \& Xu, J. H.
Electronically driven instabilities and superconductivity in the
layered La$_{2-x}$Ba$_x$CuO$_4$ perovskites. {\it Phys. Rev. Lett.}
{\bf 58}, 1035--1037 (1987).

\noindent {\it S2.}
Mattheiss, L. F.
Electronic band properties and superconductivity in La$_{2-y}$X$_y$CuO$_4$.
{\it Phys. Rev. Lett.} {\bf 58}, 1028--1030 (1987).

\noindent {\it S3.}
Perry, J. K., Tahir-Kheli, J. \& Goddard, W. A.
(unpublished) (2001).

\noindent {\it S4.}
Perdew, J. P. \& Levy, M.
Physical content of the exact Kohn-Sham orbital energies - band-gaps
and derivative discontinuities.
{\it Phys. Rev. Lett.} {\bf 51}, 1884--1887 (1983).

\noindent {\it S5.}
Sham, L. J. \& Schluter, M.
Density-functional theory of the energy-gap.
{\it Phys. Rev. Lett.} {\bf 51}, 1888--1891 (1983).

\noindent {\it S6.}
Perdew, J. P. \& Zunger, A.
Self-interaction correction to density-functional approximations
for many-electron systems.
{\it Phys. Rev. B} {\bf 23}, 5048--5079 (1981).

\noindent {\it S7.}
Svane, A.
Electronic-structure of La$_2$CuO$_4$ in the self-interaction-corrected
density functional formalism.
{\it Phys. Rev. Lett.} {\bf 68}, 1900--1903 (1992).

\noindent {\it S8.}
Temmerman, W. M., Szotek, Z. \& Winter, H.
Self-interaction-corrected electronic-structure of La$_2$CuO$_4$.
{\it Phys. Rev. B} {\bf 47}, 11533--11536 (1993).

\noindent {\it S9.}
Czyzyk, M. T. \& Sawatzky, G. A.
Local-density functional and on-site correlations - the electronic-structure
of La$_2$CuO$_4$ and LaCuO$_3$.
{\it Phys. Rev. B} {\bf 49}, 14211--14228 (1994).

\noindent {\it S10.}
Perry, J. K., Tahir-Kheli, J. \& Goddard, W. A.
Antiferromagnetic band structure of La$_2$CuO$_4$: Becke-3-Lee-Yang-Parr
calculations.
{\it Phys. Rev. B} {\bf 63}, 144510 (2001).

\noindent {\it S11.}
Xu, X. \& Goddard, W. A.
The X3LYP extended density functional for accurate descriptions of
nonbond interactions, spin states, and thermochemical properties.
{\it Proc. Nat. Acad. Sci.} {\bf 101}, 2673--2677 (2004).

\noindent {\it S12.}
Bryantsev, V. S., Diallo, M. S., van Duin, A. C. T. \& Goddard, W. A.
Evaluation of B3LYP, X3LYP, and M06-class density functionals for
predicting the binding energies of neutral, protonated, and
deprotonated water clusters.
{\it J. Chem. Theory and Comp.} {\bf 5}, 1016--1026 (2009).

\noindent {\it S13.}
Curtiss, L. A., Raghavachari, K., Trucks, G. W. \& Pople, J. A.
Gaussian-2 theory for molecular-energies of 1st-row and 2nd-row compounds.
{\it J. Chem. Phys.} {\bf 94}, 7221--7230 (1991).

\noindent {\it S14.}
Curtiss, L.~A., Raghavachari, K., Redfern, P. C. \& Pople, J. A.
Assessment of Gaussian-2 and density functional theories for the
computation of enthalpies of formation.
{\it J. Chem. Phys.} {\bf 106}, 1063--1079 (1997).

\noindent {\it S15.}
Matsuda, Y., Tahir-Kheli, J. \& Goddard, W. A.
Definitive band gaps for single-wall carbon nanotubes.
{\it J. Phys. Chem. Lett.} {\bf 1}, 2946--2950 (2010).

\noindent {\it S16.}
Xiao, H., Tahir-Kheli, J. \& Goddard, W. A.
Accurate band gaps for semiconductors from density functional theory.
{\it J. Phys. Chem. Lett.} {\bf 2}, 212--217 (2011).

\noindent {\it S17.}
Mori-Sanchez, P., Cohen, A. J. \& Yang, W. T.
Localization and delocalization errors in density functional theory and
implications for band-gap prediction.
{\it Phys. Rev. Lett.} {\bf 100}, 146401 (2008).

\noindent {\it S18.}
Haskel, D., Stern, E. A., Hinks, D. G., Mitchell, A. W. \& Jorgensen, J. D.
Altered Sr environment in La$_{2-x}$Sr$_x$CuO$_4$.
{\it Phys. Rev. B} {\bf 56}, R521--R524 (1997).
 
\noindent {\it S19.}
Shoenberg, D.
{\it Magnetic Oscillations in Metals}
(Cambridge University Press, Cambridge, UK, 1984).

\noindent {\it S20.}
Ziman, J. M.
A theory of the electrical properties of liquid metals - the monovalent metals.
{\it Phil. Mag.} {\bf 6}, 1013--1034 (1961).

\noindent {\it S21.}
Faber, T. E. \& Ziman, J. M.
A theory of electrical properties of liquid metals .3. resistivity of
binary alloys.
{\it Phil. Mag.} {\bf 11}, 153 (1965).

\noindent {\it S22.}
Nagel, S. R.
Temperature-dependence of resistivity in metallic glasses.
{\it Phys. Rev. B} {\bf 16}, 1694--1698 (1977).

\noindent {\it S23.}
Frobose, K. \& Jackle, J.
Temperature-dependence of electrical-resistivity of amorphous metals.
{\it J. Phys. F} {\bf 7}, 2331--2348 (1977).

\noindent {\it S24.}
Cote, P. J. \& Meisel, L. V.
Origin of saturation effects in electron-transport.
{\it Phys. Rev. Lett.} {\bf 40}, 1586--1589 (1978).

\noindent {\it S25.}
Meisel, L. V. \& Cote, P. J.
Structure factors in amorphous and disordered harmonic Debye solids.
{\it Phys. Rev. B} {\bf 16}, 2978--2980 (1977).


\pagebreak

\begin{figure}[htb]
\centering \includegraphics[width=0.7\linewidth]{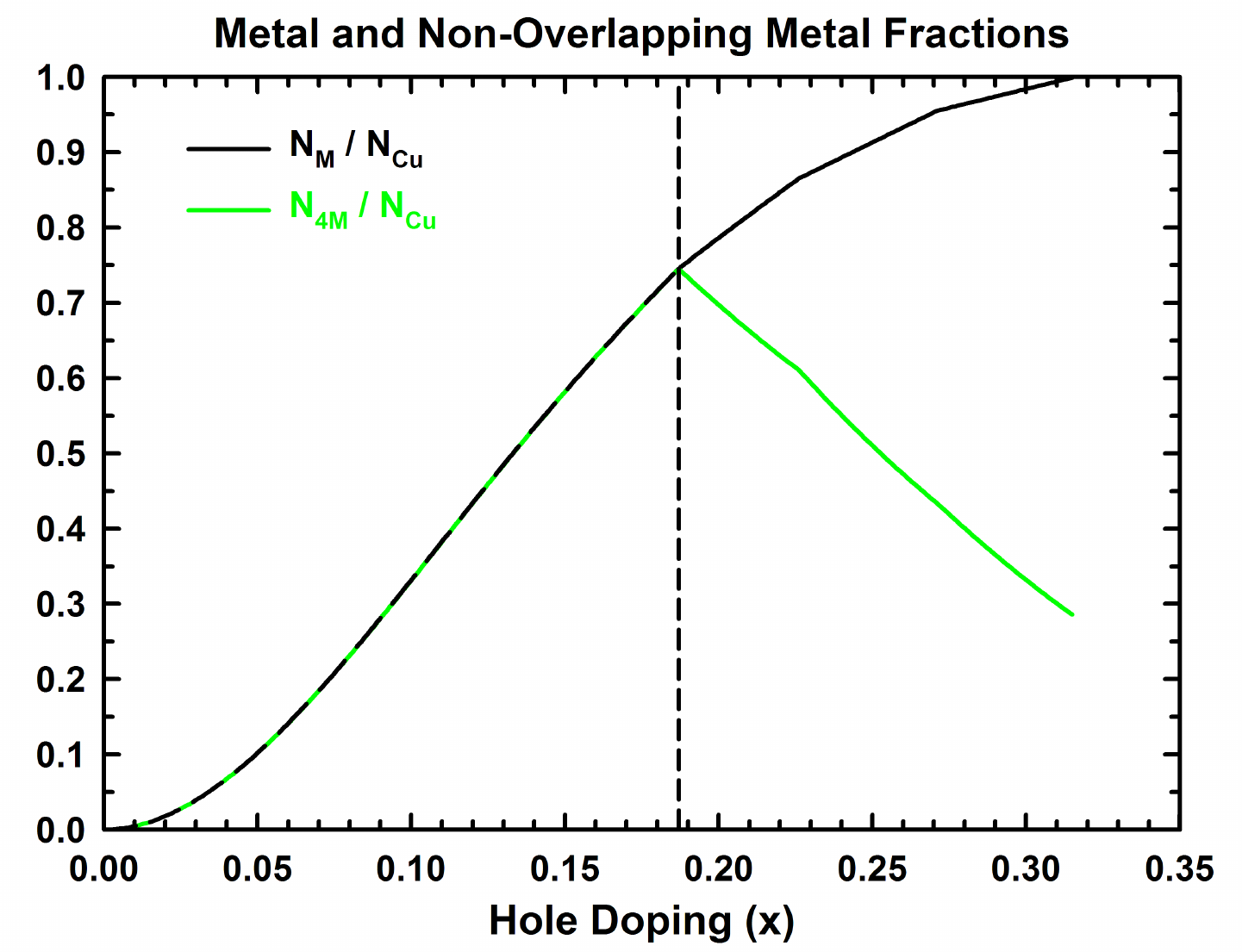}
\caption{
\textbf{Supplementary Material (FIG. S1):}
Calculated $N_{M}$ and $N_{4M}$ as a fraction of the
total number of Cu sites, $N_{Cu}$.
The green curve is
exactly the same as $\alpha_1(0)/\alpha_1(\infty)$ in figure 1.
The error bars in this figure are smaller than the width of the lines.
The algorithm used to compute these curves is described in the Methods
section.
}
\end{figure}

\pagebreak

\begin{figure}[tbp]
\centering \includegraphics[width=0.7\linewidth]{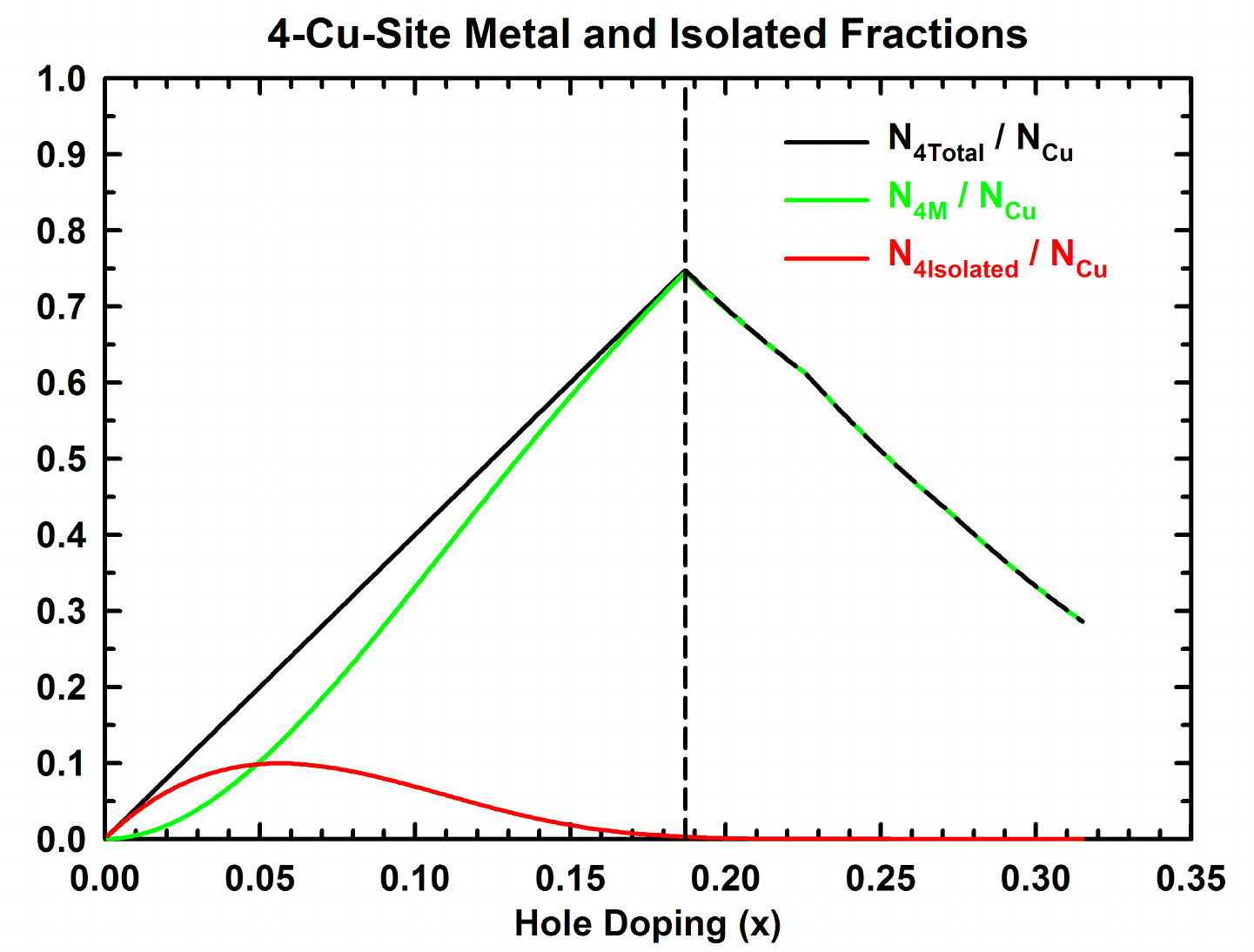}
\caption{
\textbf{Supplementary Material (FIG. S2):}
Doping evolution of the metallic and isolated
non-overlapping plaquettes. The green line is the number of Cu sites
contained in non-overlapping plaquettes
inside the metallic region per planar Cu, $N_{4M}/N_{Cu}$,
and the red line is the
number of Cu sites inside the isolated
plaquettes per planar Cu, $N_{4Isolated}/N_{Cu}$,
appearing as red squares in figure 2.
The green line is
equal to $\alpha_1(0)/\alpha_1(\infty)$ in figure 1
and is also plotted in S1.
The error bars in this figure are smaller than the width of the lines.
The isolated plaquettes are responsible for
the pseudogap \cite{Tahir-Kheli2011} and are not
relevant for the resistivity.  The black line is the total number of Cu
sites in non-overlapping plaquettes,
$N_{4Total}=N_{4M}+N_{4Isolated}$.
For $x<0.187$, there is no plaquette overlap leading to
$N_{4Total}/N_{Cu}=4x$. For $x>0.187$, $N_{4Isolated}\rightarrow 0$
very rapidly, leading to $N_{4M}\rightarrow N_{4Total}$.
}
\end{figure}

\pagebreak

\begin{figure}[htb]
\centering \includegraphics[width=0.5\linewidth]{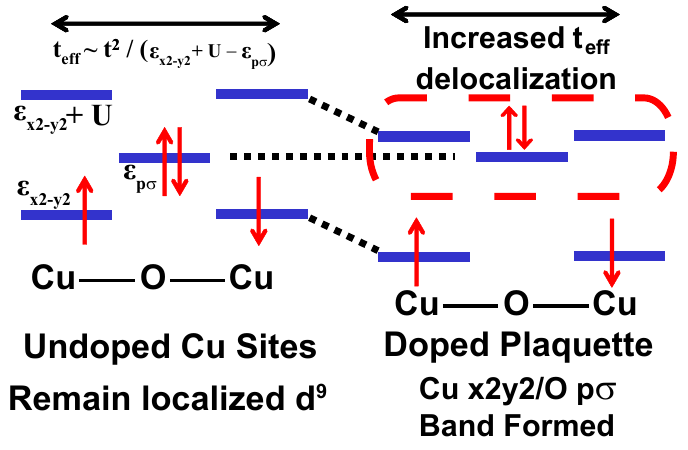}
\caption{
\textbf{Supplementary Material (FIG. S3):}
Delocalization of Cu \dxxyy/O \psigma electrons inside
4-Cu-site doped plaquette \cite{Perry2002,Tahir-Kheli2011}.
The additional hole induced by
substituting Sr for La does not go into the planar
Cu \dxxyy/O \psigma\ orbitals as is usually assumed.
Instead, it is comprised of predominantly apical O $p_z$ and Cu $d_{z^2}$
as seen in figure 3 of the main text.
The planar Cu sites surrounding the Sr dopant are
called doped sites.  Planar Cu sites that are not in the vicinity of a
dopant (only La nearby) are called undoped sites.
The energy level diagram
shows how an out-of-plane hole above the planar Cu sites leads to
delocalization of the d$^9$ Cu spins and the formation of
delocalized Cu \dxxyy/O \psigma\ states inside the plaquette.
The left figure shows the energy levels that occur at undoped Cu sites
where there are no holes in out-of-the-plane orbitals.  In this case,
the difference in energy between the doubly occupied Cu site and the
planar O orbital energy, $\epsilon_{x^2-y^2}+U-\epsilon_p$, is large,
leading to localization of spins on the Cu.  In the right figure,
the Cu orbital energy is reduced due to the missing electron in the apical
O sites directly above the Cu atoms leading to the neighboring
doubly-occupied O \psigma\ electrons delocalizing onto the Cu sites.
When two plaquettes are neighbors, the delocalization occurs over all
eight Cu sites.  When the doping is large enough that the
plaquettes percolate in $3D$ through the crystal, a ``metallic''
band is formed in the percolating swath and current can flow from one
end of the crystal to the other. This metallic band
carries the current in the normal state and becomes superconducting
below \tc.
}
\end{figure}

\pagebreak

\begin{figure}[htb]
\centering \includegraphics[width=0.8\linewidth]{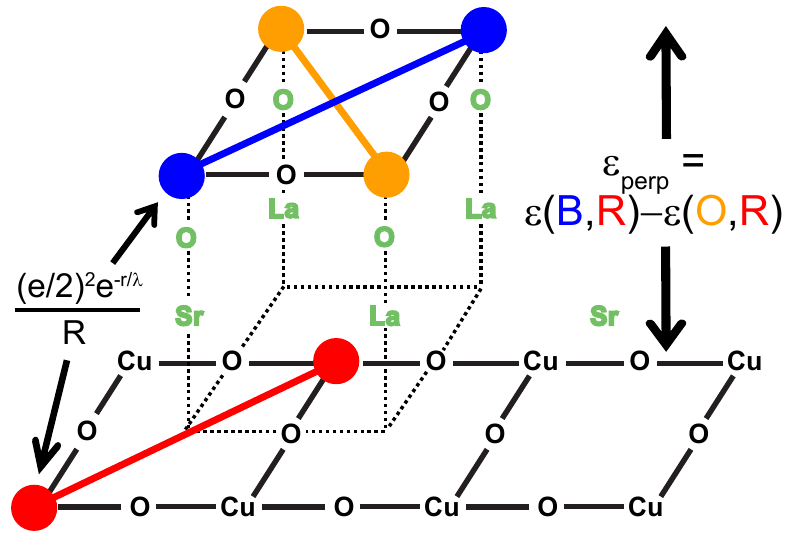}
\caption{
\textbf{Supplementary Material (FIG. S4):}
Estimating the coupling between dynamic Jahn-Teller modes in
neighboring CuO$_2$ planes. The coupling,
$\epsilon_{\mathrm{perp}}$,
is given by the difference in Coulomb repulsion energy between
the blue and red ``dumbbells", $\epsilon(B,R)$, and the
orange and red ``dumbbells", $\epsilon(O,R)$.  The $1/R$ Coulomb
repulsion is attenuated by the Thomas-Fermi metallic screening
length (dervied in the text to be, $\lambda\approx 0.69\ \mbox{\AA}$).
Since the charge in a dumbbell is divided equally between the two
ends, a charge of $(1/2)e$ is used for each interaction term
($e$ is the electron charge).  The Cu-Cu lattice spacing in
the plane is $3.8\ \mbox{\AA}$ and the distance between planes
is $6.0\ \mbox{\AA}$.
We find $\epsilon_{\mathrm{perp}}=3.6\times 10^{-5}$ eV $=0.42$ K.
}
\end{figure}

\pagebreak

\begin{figure}[htb]
\centering \includegraphics[width=0.6\linewidth]{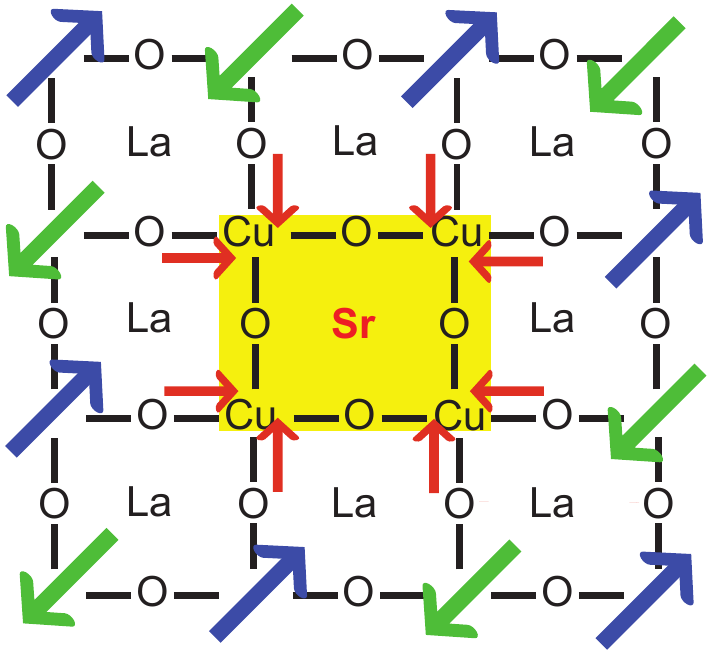}
\caption{
\textbf{Supplementary Material (FIG. S5):}
Oxygen phonon mode around an isolated plaquette.  The
figure zooms into the $4\times 4$ dotted black square in
the lower right corner of the $x=0.19$ doping in figure 2.
The yellow here denotes the isolated plaquette (shown by a red square
in figure 2).  Yellow is used here to signify that there is delocalization
of the Cu/O \dxxyy\psigma\ orbitals inside the plaquette.  Since
this plaquette is not connected to any other plaquette, these
electronic states remain localized inside the plaquette. An orbital
degeneracy at the Fermi energy of these electronic states leads to
the pseudogap \cite{Tahir-Kheli2011}.
The O atoms that reside between the localized spins in the AF
region and the delocalized electronic states inside the plaquettes
will distort and the energy of this mode will change.
We show the phonon mode by red arrows.
}
\end{figure}

\end{document}